# A Two Step Approach for Whole Slide Image Registration

Satoshi Kondo[1][0000-0002-4941-4920], Satoshi Kasai[2], Kousuke Hirasawa[3]

[1] Muroran Institute of Technology, Hokkaido, Japan
[2] Niigata University of Health and Welfare, Niigata, Japan
[3] Konica Minolta, Inc., Osaka, Japan
`kondo@mmm.muroran-it.ac.jp`

**Abstract.** Multi-stain whole-slide-image (WSI) registration is an active field of research. It is unclear, however, how the current WSI registration methods would perform on a real-world data set. AutomatiC Registration Of Breast cAncer Tissue (ACROBAT) challenge is held to verify the performance of the current WSI registration methods by using a new dataset that originates from routine diagnostics to assess real-world applicability. In this report, we present our solution for the ACROBAT challenge. We employ a two-step approach including rigid and non-rigid transforms. The experimental results show that the median 90th percentile is 1,250 um for the validation dataset.

## 1 Introduction

Multi-stain whole-slide-image (WSI) registration is an active field of research and has previously been addressed in the ANHIR challenge [1]. It is unclear, however, how the current WSI registration methods would perform on a real-world data set from routine clinical workflows, where artifacts such as glass cracks, pen marks or air bubbles are common.

AutomatiC Registration Of Breast cAncer Tissue (ACROBAT) challenge [2] is held to verify the performance of the current WSI registration methods by using a new dataset that originates from routine diagnostics to assess real-world applicability. The training data set will contain WSIs from 750 cases, with one H&E WSI per case which is a fixed image and up to four IHC WSIs out of four routine diagnostic stains (H&E, ER, PR, HER2, KI67) which are moving images in image registration. Algorithms can be tuned with a validation data set with 100 cases and evaluated with a test set consisting of 300 cases.

In this report, we present our solution to the ACROBAT challenge. We employ a two-step approach to handle both large translation and non-rigid transform.

## 2 Proposed Method

Figure 1 shows an example of paired WSI. These pictures are obtained from the same sample and the staining methods are different. As can be seen from Fig. 1, there is large translation between these images. There can be seen about 180 degree rotation between some paired WSI.

Figure 2 shows an overview of our proposed method. Our method is a two-step approach to handle both large translation and non-rigid transform. A fixed image, H&E image, and a moving image, one of IHC images, are converted to gray scale images and applied some preprocessing at first. As can be seen in Fig. 1(a), some images have black area around the border of the image, and we remove this area by checking the pixel values in the border area in the preprocessing.

At the first step of the registration, our method estimates translation (in horizontal and vertical directions) and rotation (0 or 180 degree) between the fixed and moving images. We employ a template matching method to estimate translation and rotation. The matching criteria is normalized cross correlation. The template matching is performed by using the original shape moving image and the 180-degree rotated moving image.

At the second step of the registration, our method performs non-rigid registration between the fixed image and the translated / rotated moving image. We employ VoxelMorph [3]. In VoxelMorph, registration task is formulated as a function that maps an input image pair to a deformation field that aligns these images. The function is parameterized via a convolutional neural network (CNN), and the parameters of the neural network are optimized on a set of images.

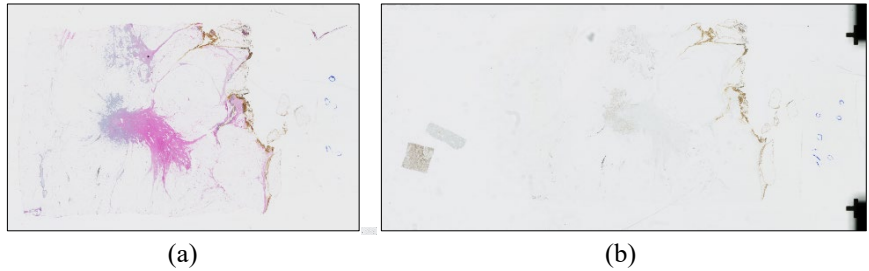

(a)                                            (b)

Figure 1. An example of paired WSI. (a) H&E image. (b) ER image.

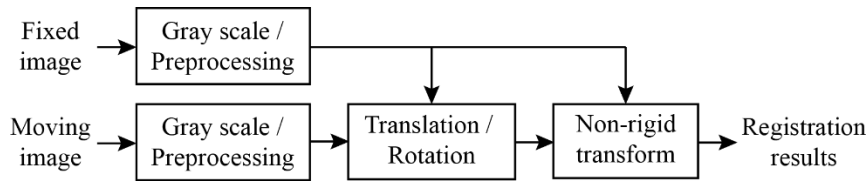

Figure 2. Overview of our proposed method.

## 3   Experiments

We evaluated our proposed method by using dataset provided by ACROBAT organizers. All images are WSIs generated by a Hamamatsu Nanozoomer XR at 0.23 µm per pixel (40X). The numbers of the training and validation data are 2,656 and 100 fixed/moving image pairs. For registration, we down-sample the original (40X) image at 1/32 scale.

The first stage, i.e. template matching, is performed with pure image processing. We do not use any machine learning technique for the template matching. In the second stage, i.e., VoxelMorph, we use U-Net type deep learning network which is the same architecture proposed in the original article [3]. In the training of VoxelMorph, the initial learning rate is 0.001 changing with cosine annealing, the batch size is 8, the number of epochs is 90. We use MSE loss and do not use bidirectional loss. The optimizer is Adam [4]. We use blurring, shift, scaling, and rotation as augmentation.

We evaluated our method with the evaluation system provided by the organizers of ACROBAT. In the evaluation, a set of manually generated landmarks images are used. The distance between registered and target landmark is computed for annotations. The main evaluation metric is the median 90th percentile across the distances between all paired landmarks in an image. The evaluation result for our method is 1,205 um.

## 4   Conclusions

In this report, we presented our solution for the ACROBAT challenge. We employ a two-step approach including rigid and non-rigid transforms. The experimental results show that the median 90$^{th}$ percentile is 1,250 um for the validation dataset.